\documentclass[lettersize,journal]{IEEEtran}
\IEEEoverridecommandlockouts
\usepackage{float}
\usepackage{threeparttable}
\usepackage{multirow}
\usepackage{makecell}
\usepackage{cite}
\usepackage{amsmath,amsfonts}
\usepackage{array}
\usepackage{booktabs} 
\usepackage[caption=false,font=normalsize,labelfont=sf,textfont=sf]{subfig}
\usepackage{textcomp}
\usepackage{stfloats}
\usepackage{url}
\usepackage{verbatim}
\usepackage{graphicx}
\usepackage{algpseudocode}
\usepackage{caption}
\usepackage{amsmath}
\usepackage[ruled,vlined]{algorithm2e}
\usepackage{algpseudocode} 
\usepackage{caption}
\usepackage{xcolor}
\usepackage{float}

\usepackage{tabularx}
\usepackage{colortbl}
\definecolor{mygray}{gray}{.9}
\definecolor{mygray1}{gray}{.8}
\definecolor{mygray2}{gray}{.7}
\definecolor{mygray3}{gray}{.6}
\usepackage{diagbox}
\usepackage{enumitem}
\usepackage{tikz}

\usepackage[switch]{lineno,}
\usepackage{xcolor}

\newcommand*{\circledd}[1]{\lower.7ex\hbox{\tikz\draw (0pt, 0pt)
   circle (.4em) node {\makebox[0.25em][c]{\small#1}};}}

\newcommand*\circled[1]{\tikz[baseline=(char.base)]{
       \node[shape=circle,fill,inner sep=1pt] (char) {\textcolor{white}{\small#1}};}}

\usepackage{xcolor}
\definecolor{mygray}{gray}{.9}

\begin{document}

\title{

pHNSW: PCA-Based Filtering to Accelerate HNSW Approximate Nearest Neighbor Search}
\author{~\IEEEmembership{}

{Zheng Li$^1$}\hspace{0.2cm}
{Guangyi Zeng$^1$}\hspace{0.2cm}
{Paul Delestrac$^2$}\hspace{0.2cm}
{Enyi Yao$^1$}\hspace{0.2cm}
{Simei Yang$^{1*}$}
\vspace{-0.1cm}
\\
\small $^1$South China University of Technology, Guangzhou, China\hspace{0.5cm}
$^2$KTH Royal Institute of Technology, Stockholm, Sweden
\vspace{0.1cm}
\\
\small $^*$Corresponding Email: yangsimei@scut.edu.cn
\vspace{-0.6cm}

 \thanks{
 Permission to make digital or hard copies of all or part of this work for personal or classroom use is granted without fee provided that copies are not made or distributed for profit or commercial advantage and that copies bear this notice and the full citation on the first page. Copyrights for components of this work owned by others than ACM must be honored. Abstracting with credit is permitted. To copy otherwise, or republish, to post on servers or to redistribute to lists, requires prior specific permission and/or a fee. Request permissions from permissions@acm.org.\\
  ASP-DAC ’26, January 13–16, 2026, Hong Kong, China\\
  © 2026 Association for Computing Machinery. 
  \\
  \\
  }

}

\markboth{}%
{Shell \MakeLowercase{\textit{et al.}}: A Sample Article Using IEEEtran.cls for IEEE Journals}     


\maketitle

\makeatletter
\markboth{Accepted for publication in the Proceedings of ASP-DAC 2026}{Accepted for publication in the Proceedings of ASP-DAC 2026}
\thispagestyle{headings}
\makeatother


\begin{abstract}

Hierarchical Navigable Small World (HNSW) has demonstrated impressive accuracy and low latency for high-dimensional nearest neighbor searches. However, its high computational demands and irregular, large-volume data access patterns present significant challenges to search efficiency. To address these challenges, we introduce pHNSW,  an algorithm-hardware co-optimized solution that accelerates HNSW through Principal Component Analysis (PCA) filtering. On the algorithm side, we apply PCA filtering to reduce the dimensionality of the dataset, thereby lowering the volume of neighbor access and decreasing the computational load for distance calculations. On the hardware side, we design the pHNSW processor with custom instructions to optimize search throughput and energy efficiency. In the experiments, we synthesized the pHNSW processor RTL design with a 65nm technology node and evaluated it using DDR4 and HBM1.0 DRAM standards. The results show that pHNSW boosts Queries per Second (QPS) by 14.47×$\sim$21.37× on a CPU and 5.37×$\sim$8.46× on a GPU, while reducing energy consumption by up to 57.4\% compared to standard HNSW implementation.

\end{abstract}


\section{Introduction}
\label{intro}

Approximate Nearest Neighbor (ANN) search is a fundamental technique widely used in artificial intelligence and machine learning, aiming to efficiently identify the closest feature vectors to a given query in high-dimensional datasets. It plays a critical role in large-scale data retrieval tasks, including information retrieval, recommendation systems, and other applications involving vector similarity search~\cite{yin2025panns}. Hierarchical Navigable Small World (HNSW) ~\cite{HNSW} is a graph-based ANN algorithm, which can achieve top performance (e.g., speed, accuracy, scalability) in ANN-benchmarks~\cite{benchmark}. Its high search efficiency primarily comes from its hierarchical graph structure with multiple graph layers, where each forms a small-world graph that captures varying levels of data proximity.

However, HNSW faces significant computational and memory challenges due to the demands of high-dimensional data processing. For instance, in the ANN benchmark SIFT1M~\cite{jegou2010product}, each 128-dimensional vector occupies 512 Bytes. The HNSW algorithm requires intensive distance computations and sorting operations, with computational costs growing rapidly due to increasing data dimensionality or dataset size. Additionally, it demands extensive off-chip memory access, since each graph node can have multiple neighbors across various layers.
Moreover, efficient HNSW traversal demands irregular memory access patterns, as graph structures inherently lack a predictable, contiguous layout in memory.

To address HNSW's computational and memory bottlenecks, prior works focus on optimizing computational parallelism through specialized hardware units for distance computation and sorting~\cite{Peng, Kim}, and on improving off-chip data access bandwidth using processing-in-memory (PIM) architectures~\cite{zhu2023processing, wang2024ndsearch}. Orthogonal to prior works, our work adopts a different strategy by prioritizing data dimensionality reduction based on the Principal Component Analysis (PCA) filtering~\cite{abdi2010principal} to improve search efficiency. PCA is a technique that reduces a dataset’s dimensionality by projecting it into a lower-dimensional space via an orthogonal transformation~\cite{abdi2010principal}. This process preserves the most significant features by capturing the maximum variance within the dataset. The studies in~\cite{PCAF,song2022accelerating} explore the use of PCA filtering for the K-Nearest Neighbors (KNN) problem. These studies first reduce dataset dimensionality through PCA transformation, then project the filtered top-k candidates back into the original high-dimensional space to preserve search accuracy. We consider~\cite{song2022accelerating,PCAF}, denoted as pKNN, the most closely related works to ours.

In this paper, we integrate PCA filtering into HNSW and present \textit{pHNSW}, an algorithm-hardware co-optimization design to accelerate graph search. We summarize the \textbf{main contributions} and  \textbf{novelties} of this work as follows:

\begin{itemize}
    \item
    
    \textit{Algorithm Optimization}: We propose the pHNSW algorithm, which optimizes HNSW based on PCA filtering to reduce data dimensionality and introduces a $k$-value selection method to balance accuracy and performance. Unlike pKNN~\cite{song2022accelerating, PCAF}, which uses a single $k$-value, pHNSW employs a hierarchical graph structure where the $k$-value varies across different hierarchical layers.

    \item \textit{DataSbase Organization}: We optimize the database structure to eliminate irregular data access, enabling parallel processing of computations and sorting. In contrast, pKNN~\cite{song2022accelerating} performs distance calculations for low-dimensional data sequentially with irregular data access.

    \item \textit{Hardware Design}: We design the pHNSW processor with a custom instruction set architecture (ISA) and a dedicated search module to optimize search throughput and energy efficiency, whereas pKNN~\cite{song2022accelerating} utilizes High-Level Synthesis (HLS), which typically delivers lower performance compared to a custom hardware design.

    \item We synthesize the RTL of the pHNSW processor using a 65nm technology node and evaluate search throughput (QPS) and energy consumption with DDR4 and HBM1.0 DRAM standards. Experiments on SIFT1M show up to 21.37× higher QPS and 57.4\% lower energy consumption than standard HNSW implementation.

\end{itemize}

The remainder of the paper is organized as follows.  Section~\ref{relatedwork} discusses the related work. Then, Section~\ref{algo} introduces the proposed pHNSW algorithm. Section~\ref{sec:hw} details the optimized database organization and the pHNSW processor design. Section~\ref{sec:Exp} provides experimental evaluations of the processor's area, search performance, and energy consumption. Finally, Section~\ref{sec:conclusion} concludes the paper and outlines potential directions for future work.

\section{Related Work}
\label{relatedwork}

This section reviews existing work on addressing the computational and memory bottlenecks of graph-based ANNs, with a particular emphasis on HNSW.

\begin{table}[h]
\caption{Summary of State-of-the-Art Research}
\renewcommand\arraystretch{1.5}
\scriptsize
\begin{center}
\begin{threeparttable}
\vspace{-0.5cm}
\begin{tabular}{|m{2cm}<{\centering}|m{2.6cm}<{\centering}|m{3cm}<{\centering}|}\hline

\bf Paper-Year & \bf Algorithm Optimization &  \bf Hardware Optimization  \cr\hline

HNSW\cite{HNSW}-2018&Propose  HNSW (C+S)& CPU Implementation\cr\hline
\cite{GGNN}-2023;\cite{CAGRA}-2024&\multirow{1}{*}{Propose GB-ANN (C+S)} & GPU batch processing; Data structure optimization\cr\hline
\cite{tian2025towards}-2025& Consider GB-ANN (S) & CPU+GPU collaboration\cr\hline\hline

\cite{Peng}-2021;\cite{Kim}-2022   & \multirow{2}{*}{/}
 &HNSW-ACC (S) \cr\cline{1-1}\cline{3-3}

\cite{zhu2023processing}-2023;\cite{wang2024ndsearch}-2024 &&GB-ANN-ACC (S), PIM \cr\hline

\cellcolor{mygray}pKNN\cite{song2022accelerating}-2022 &PCA Filtering KNN~\cite{PCAF}&KNN-ACC(S), HLS\cr\hline

\cellcolor{mygray}Our work &Optimize HNSW with PCA filtering (S)&HNSW-ACC (S) with custom ISA, Database Optimization\cr\hline\hline

Flash\cite{wang2025accelerating}-2025& Data preprocessing for HNSW (C) &CPU SIMD implementation\cr\hline

\end{tabular}
    \begin{tablenotes}
    \footnotesize
    \item *Note: \textbf{C} and \textbf{S} represent the graph construction and query search phases, respectively; \textbf{GB-ANN} refers to graph-based ANNs; \textbf{ACC} stands for accelerator design; \textbf{PIM} denotes Processing-In-Memory architecture.
    \end{tablenotes}
    \end{threeparttable}
    \vspace{-0.35cm}
\label{table:RelatedWork}
\end{center}
\end{table}

Table~\ref{table:RelatedWork} summarizes state-of-the-art research on both algorithmic and hardware-level optimizations. The work in~\cite{HNSW} introduces the HNSW algorithm, which, similar to other graph-based ANN approaches, consists of two main phases: \textit{graph construction (C)} and \textit{query search (S)}. The \textit{C} phase constructs a multi-layered graph from the dataset, while the \textit{S} phase searches for the approximate nearest neighbors of a given query. The original HNSW algorithm~\cite{HNSW} is implemented on a CPU. The studies in~\cite{GGNN, CAGRA} exploit the parallelism of graph-based ANNs on GPUs through batch processing and develop GPU-friendly data structures to optimize memory access efficiency in both \textit{C} and \textit{S} phases. The work in~\cite{tian2025towards} introduces a CPU/GPU collaborative filtering strategy, where the CPU handles initial vector selection and sends only the selective query vector IDs to the GPU. This approach eliminates the need to transfer high-dimensional raw data to GPU, thereby reducing data transfer overhead in \textit{S} phase.

To optimize both search performance and energy efficiency, the works of~\cite{Peng, Kim, zhu2023processing, wang2024ndsearch} propose hardware accelerators with specialized units to optimize data access, distance calculation, and sorting. In particular, the works in~\cite{zhu2023processing,wang2024ndsearch} introduce processing-in-memory (PIM) architectures for graph-based ANNs, integrating computation units within both main memory and Solid State Drives (SSDs) to maximize data access bandwidth. Instead of optimizing computational parallelism or memory access through dedicated hardware accelerators or PIM architectures, the works in~\cite{song2022accelerating, PCAF} focus on reducing data dimensionality to improve search efficiency. The pKNN~\cite{song2022accelerating} work implements a hardware accelerator for the PCA-based filtering algorithm for the KNN problem~\cite{PCAF}. Similarly, our algorithm–hardware co-design, \textit{pHNSW}, incorporates PCA filtering into HNSW to improve search efficiency. Compared to pKNN~\cite{song2022accelerating,PCAF}, our pHNSW design introduces three key innovations across the algorithm, database organization, and hardware architecture, as detailed in the contribution part of Section~\ref{intro}.

A recent work, Flash~\cite{wang2025accelerating}, optimizes the HNSW graph index during the \textit{C} phase by first reducing storage costs through PCA and product quantization for dimensionality compression, and then optimizing the index layout to improve memory access efficiency in CPU SIMD (Single Instruction Multiple Data) execution. 
Unlike~\cite{wang2025accelerating}, which uses PCA as a preprocessing step without modifying the HNSW algorithm, our proposed pHNSW algorithm performs the initial candidate search in a PCA-based low-dimensional space, then back-projects the top-$k$ filtered candidates to the high-dimensional space for accurate distance computation. Additionally, while~\cite{wang2025accelerating} builds the HNSW graph on a CPU, we introduce a custom pHNSW processor with a dedicated ISA to accelerate the search phase through algorithm–hardware co-design.

\section{pHNSW Algorithm}
\label{algo}

This section presents the pHNSW algorithm and discusses the selection of its key parameter \textit{filter size k}, which highly impacts memory access efficiency and search accuracy.

\subsection{pHNSW Algorithm}
\label{P-algo}

\begin{figure}[h]
    \vspace{-0.2cm}
	\centering
\includegraphics[width= 1\textwidth]{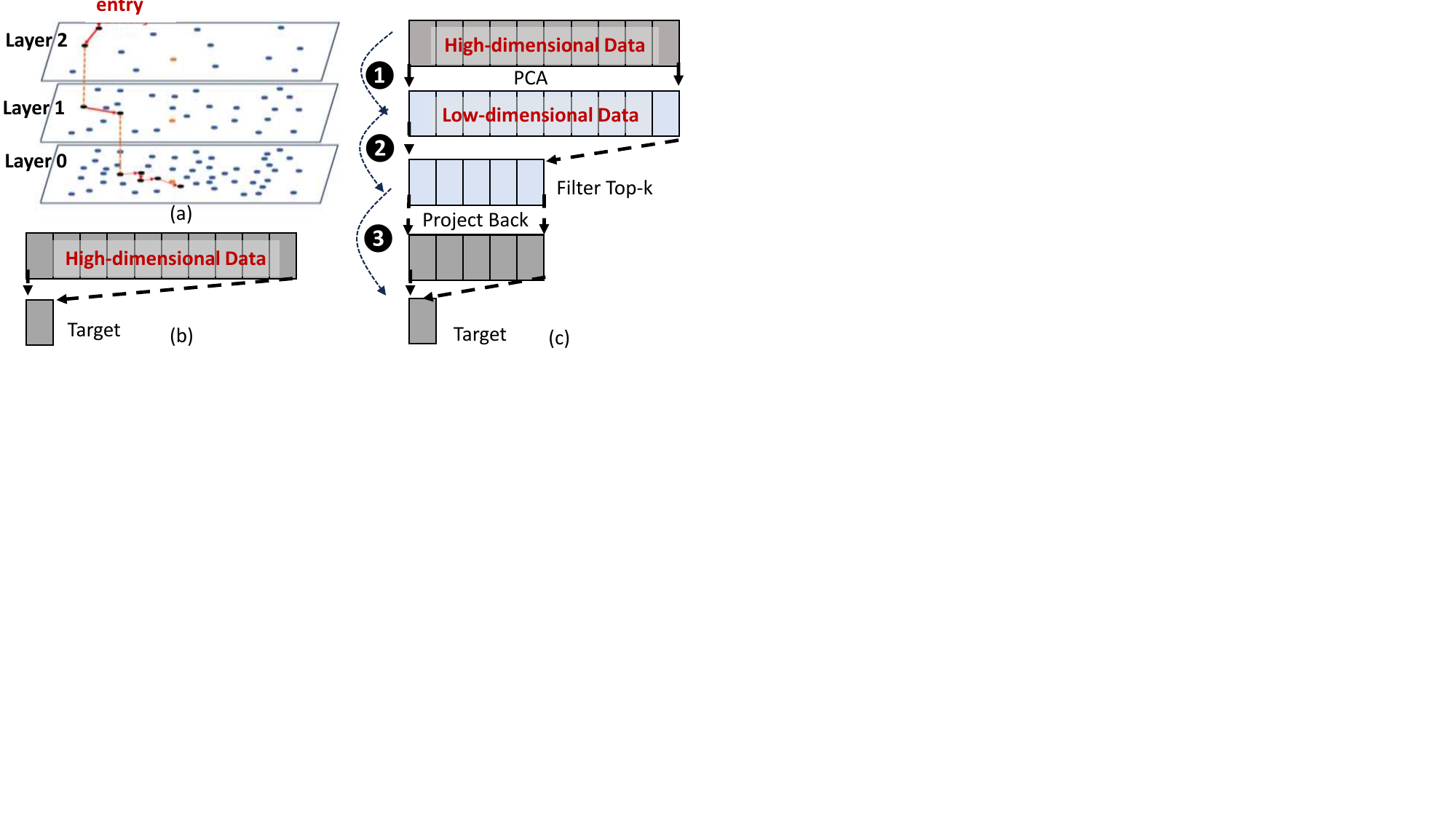}
	\vspace{-6cm}
	\caption{(a) Hierarchical graph of HNSW; (b) HNSW searches from a list of neighbors in high-dimensional space; (c) pHNSW searches based on PCA filtering in three steps.}
	\label{fig:p-hnsw}
\end{figure}

Fig.\ref{fig:p-hnsw} highlights the differences between the standard HNSW and the proposed pHNSW. Fig.\ref{fig:p-hnsw} (a) shows HNSW's hierarchical graph, where the search moves from higher layers (distant points) to lower layers (closer points) to progressively locate the nearest neighbors. Fig. \ref{fig:p-hnsw} (b) shows HNSW's goal of finding the data point(s) most similar to the query vector from a list of neighbors in high-dimensional space. Fig.~\ref{fig:p-hnsw} (c) illustrates how pHNSW leverages PCA filtering to locate target data point(s) from a list of neighbors in three steps. Step \circled{\small{\bf{1}}} applies PCA~\cite{PCAF} to transform the database from a high-dimensional space to a lower-dimensional space (e.g., for the ANN benchmark SIFT1M, from 128 to 15 dimensions, with each dimension occupying 4 Bytes). Step \circled{\small{\bf{2}}} performs distance calculations and sorting in the low-dimensional space to filter the top-$k$ candidates. Step \circled{\small{\bf{3}}} projects these top-$k$ candidates back into the high-dimensional space for precise distance computation. Compared to HNSW~\cite{HNSW}, pHNSW mainly operates in low-dimensional space and computes distances in high-dimensional space only $k$ times. 

\setlength{\textfloatsep}{5pt}
\begin{algorithm}[t]
\small
\caption{pHNSW Algorithm}
\label{Algo:P-HNSW}
\LinesNumbered
\SetKwInOut{Input}{Input} 
\SetKwInOut{Output}{Output}
\SetKwInOut{Params}{Params}

\KwIn{query $q$, enter-point $ep$, number of nearest to return $ef$, layer number $l$, low-dimensional query in  $q\_pca$}
\KwOut{$ef$ closest neighbors to $q$}
\Params{$V$: visited list, $C$: candidate list, $F$: final list, 
$C\_pca$: candidate list in low-dimensional space, $C\_pca\_tmp$: temporary heap of $C\_pca$,  $k$: filter size by PCA filtering
}

$V,C,F \gets ep$ \\
\While{$|C| > 0$} {
$c \gets$ extract nearest element from $C$ to $q$\\
$f \gets$ extract furthest element from $F$ to $q$\\
$f\_pca \gets$ get furthest element from $C\_pca$ to $q\_pca$\\
$C\_pca \gets$$\emptyset$\\
\If{$Dist(c,q) > Dist(f,q)$}{
    \textbf{Break}}

//For-loop of step~\circled{\small{\bf{2}}} in Fig~\ref{fig:p-hnsw}(c)

\For{$each$ $e \in Neighbor(c)$ at layer $l$}{
  \If{$Dist(e\_pca,q\_pca)$$ < $$Dist(f\_pca,q\_pca)$} {
    $C\_pca\gets$$e\cup$$C\_pca$}
   }
  
\textbf{\textit{Update}} $C\_pca$ to keep top-$k$ smallest element

//For-loop of step~\circled{\small{\bf{3}}} in Fig.~\ref{fig:p-hnsw}(c)

\For{$each$ $m \in$ $C\_pca$}{
    \If{$m \notin$ $V$}{
       $V \gets$ $V \cup$$m$\\
        $f \gets$ extract furthest element from $F$ to $q$\\
        \If{$Dist(m,q)$$ < $$Dist(f,q)$ or $|F|$ $<$ $ef$} {
        $C\_pca\_tmp \gets$ $C\_pca\_tmp \cup$$m$\\
        $C \gets$ $C \cup$$m$, 
        $F \gets$ $F \cup$$m$\\
        \If{$|F|$ $>$ $ef$}{
        remove furthest element from $F$              
               }}}}
\textbf{\textit{Update}} $C\_pca \gets$ $C\_pca\_tmp$ 
}
\textbf{return} $F$

\end{algorithm}

Algorithm~\ref{Algo:P-HNSW} details the pHNSW search process. We assume that all PCA-transformed low-dimensional data corresponding to the preconstructed graph is stored in off-chip memory, with the storage format detailed in Section~\ref{sub:dataStorage}. Similar to the original HNSW  in~\cite{Kim}, the pHNSW algorithm utilizes three lists, visited list $V$, candidate list $C$ and final list $F$ to find the closest neighbors. Parameters associated with the PCA transformation are denoted by the suffix $\_pca$. Given a target query $q$, the pHNSW algorithm starts the search at the top level with a randomly chosen entry point $ep$ and searches for the $ef$ nearest points in this layer (e.g., for SIFT1M, $ef=1$ for Layer 1-5, $ef=10$ for Layer 0). The search continues as long as there are remaining candidates in $C$. If the distance from $q$ to the nearest candidate in $C$ exceeds the distance from $q$ to the farthest point in $F$, pHNSW stops searching in the current layer and proceeds to the next layer (lines 7-8). Otherwise, the search proceeds to the for-loop of step~{\circled{\small{\bf{2}}} (lines 9–12), where $C\_pca$ is updated based on distance comparisons to retain the top-$k$ closest elements (line 13). This is the key distinction between pHNSW and HNSW~\cite{HNSW}, enabling most searches in a lower-dimensional space and limiting high-dimensional iterations to $k$ times in the subsequent step. In Step~{\circled{\small{\bf{3}}}} (lines 14-23), the algorithm checks whether the point has been visited (line 16) and then updates $C\_pca\_tmp$ (the temporary heap of $C\_pca$), $C$, and $F$ based on distance comparison results (lines 19–23). This process repeats until no candidates remain in the current layer, at which the elements in $F$ become the entry point for the next layer.

\subsection{Discussion on Top-k Parameter Selection}
\label{filtersize}

The \textit{filter size} $k$ (in top-$k$, step~{\circled{\small{\bf{2}}}) is a key parameter that significantly impacts both search accuracy and search throughput, typically measured by recall rate (e.g., Recall@10, the proportion of true nearest neighbors found among the top 10 results) and queries per second (QPS).
Since point density increases at lower layer levels in pHNSW (or HNSW) graphs, as previously shown in Fig.~\ref{fig:p-hnsw}, we adjust $k$ differently across layers to trade-off search performance and accuracy. Given a six-layer-graph database of SIFT1M, the hierarchical graph contains $M$ (e.g.,16) neighbors from layers 1-5, and $2M$ (e.g.,32) neighbors at layer 0. We set the $k$-values for Layers 2–5 to 3, following the recommendation in~\cite{PCAF}, which suggests setting $k$ to three times the number of search candidates ($ef = 1$). In denser layers, we set higher $k$ values due to the increased number of candidates (e.g., $ef=10$ for Layer 0). Based on our experimental evaluations in Fig.~\ref{fig:recall}, we set 
$k$ to 8 for Layer 1 and 16 for Layer 0, achieving a recall@10 of 0.92. Our experiments show that increasing 
$k$ beyond the selected values for Layers 0-1  does not significantly improve the recall rate, but it can reduce QPS by up to 21.4\% (when 
$k$ for Layer 0 is 18, Fig.~\ref{fig:recall}.b).

\begin{figure}[t]
	\centering
\includegraphics[width= 1\textwidth]{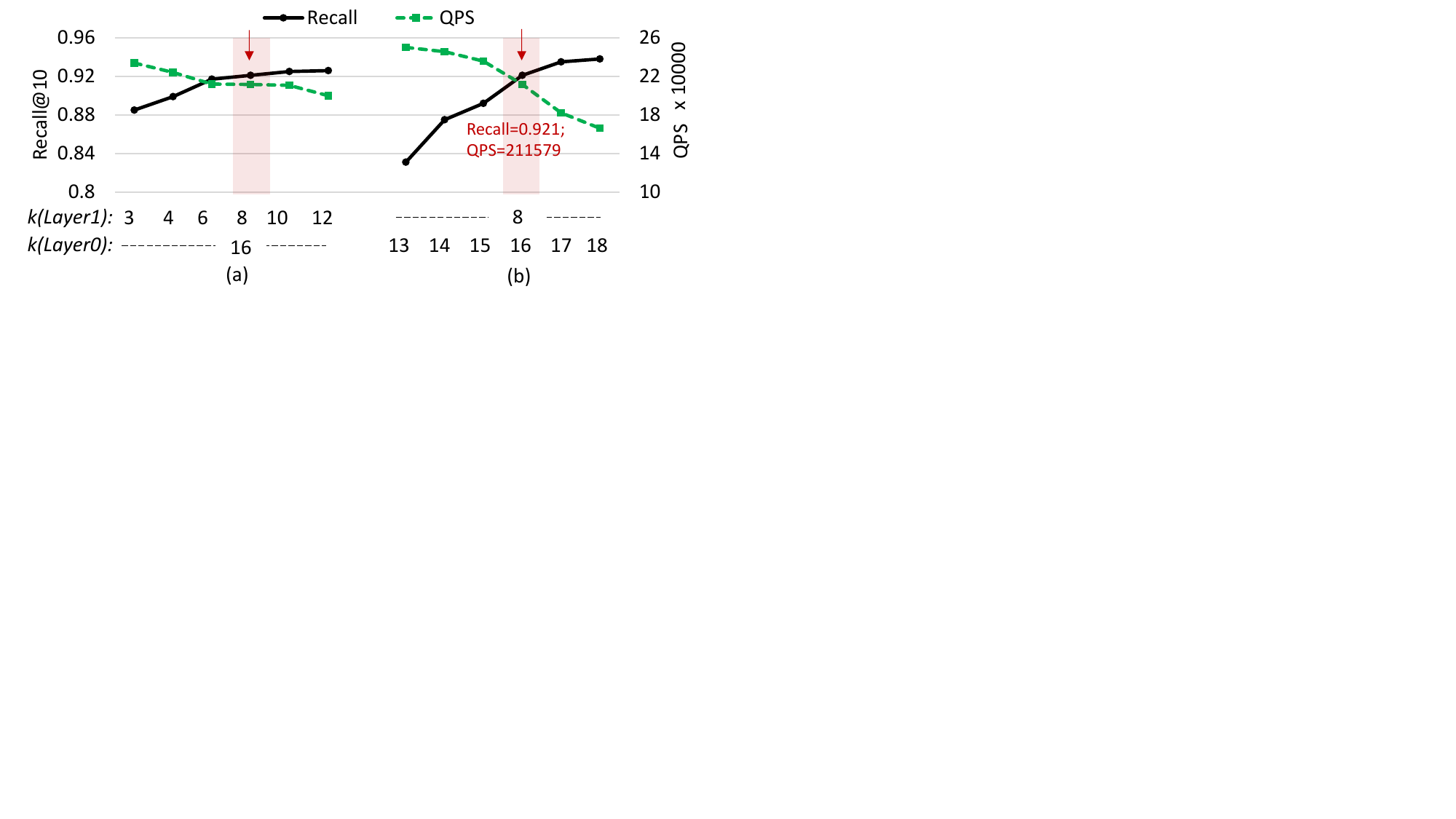}
	\vspace{-6.9cm}
\caption{Recall@10 and QPS evolutions: (a) Different k(Layer1) with k(Layer0)=16; (b) Different k(Layer0) with k(Layer1)=8.}
	\label{fig:recall}
\end{figure}

\section{pHNSW Processor}
\label{sec:hw}
\begin{figure*}[!t]
    \centering
\includegraphics[width=\textwidth]{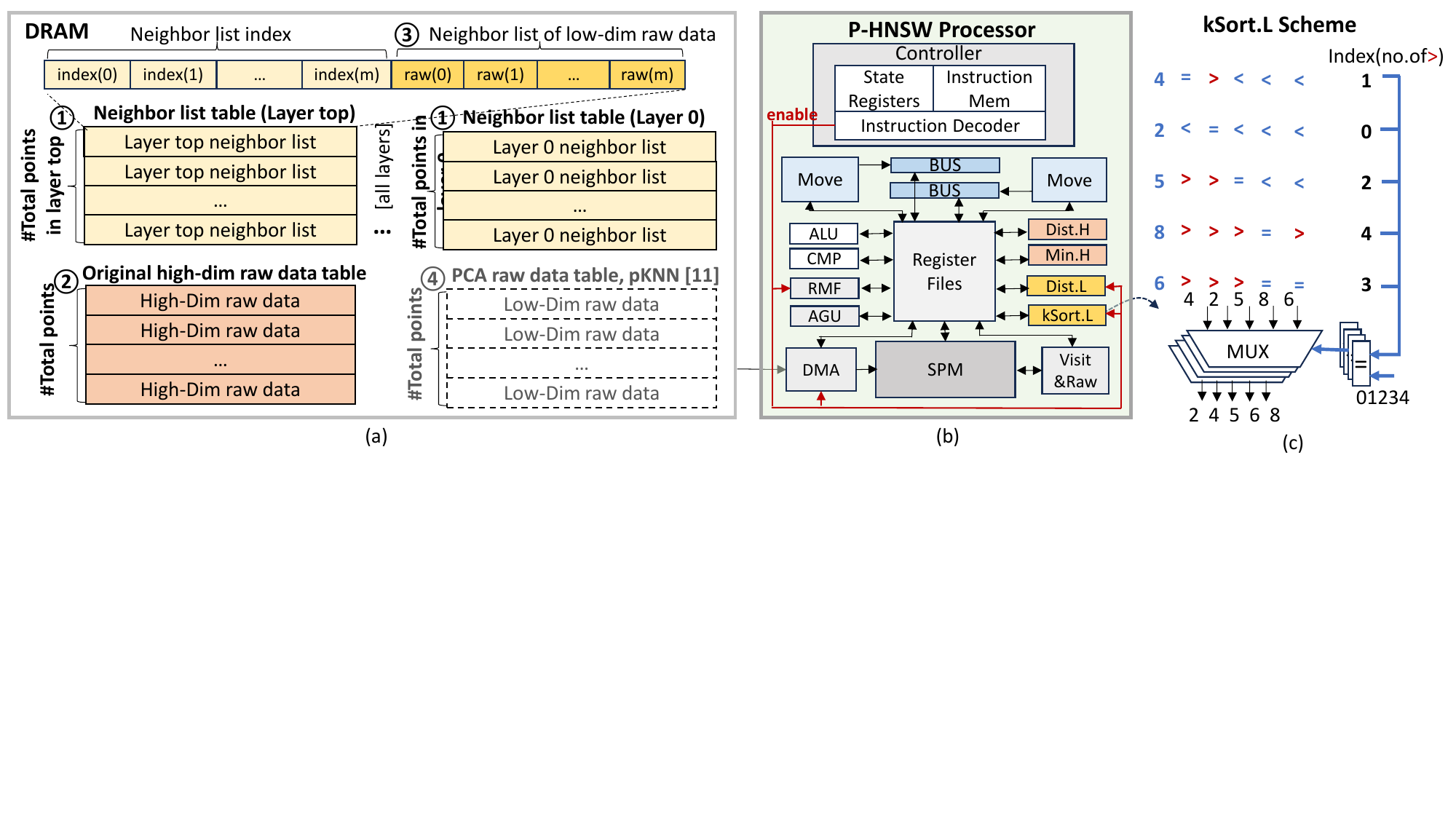} 
\vspace{-5cm}
    \caption{(a) Off-chip database organization; (b) pHNSW processor design (*Table~\ref{Table:ISA} lists the functionalities of the components); (c) The full parallel sorting scheme of the \textit{kSort.L} module, for an example of sorting five data elements.}
    \label{fig_3}
\end{figure*}

In this section, we describe the off-chip memory organization of the database, as well as the hardware design and dataflow of the custom pHNSW processor.

\subsection{Database Organization}
\label{sub:dataStorage}

Fig.~\ref{fig_3}(a) illustrates the organization of the database in off-chip memory, consisting of two types of tables: \circledd{\small{\bf{1}}} an index table for each layer, which stores lists of neighbor indexes followed by low-dimensional raw data; and \circledd{\small{\bf{2}}} an original raw data table, which contains the high-dimensional data for each point. Compared to the prior work in~\cite{Kim}, our database organization introduces additional neighbor lists for low-dimensional raw data (see \circledd{\small{\bf{3}}}). This increases memory usage by 1.8GB—approximately 2.92$\times$ the size of the SIFT1M dataset—but allows retrieving the entire neighbor list in a single access, significantly reducing irregular feature access. In contrast, pKNN~\cite{song2022accelerating} stores low-dimensional data separately (\circledd{\small{\bf{4}}} in Fig.~\ref{fig_3}(a)), resulting in irregular access due to unpredictable neighboring feature indexes. A detailed comparison of low-dimensional raw data storage is provided in Section~\ref{QPS}.

\subsection{pHNSW Processor Design}

Fig.\ref{fig_3}(b) shows an overview of the pHNSW processor, including a controller, memory units, and compute units. 

\subsubsection{Controller}

The controller handles instruction fetch, decode, and execution. Table~\ref{Table:ISA} summarizes the proposed custom instructions, each sized at 32 bits. Our design supports parallel data movement between registers via two \textit{Move} components and two \textit{BUS} units. The \textit{Move} instructions manage data movement between registers during operations like distance computation and sorting, which account for the majority of instructions executed during a search (up to 72.8\%). This dual data movement design can improve data transfer efficiency.

\subsubsection{Memory Units and Data Access}

Unlike pKNN~\cite{song2022accelerating}, which relies on irregular DRAM access for low-dimensional data, our pHNSW processor ensures regular access via optimized DRAM organization (\circledd{\small{\bf{4}}} vs. \circledd{\small{\bf{3}}} in Fig.~\ref{fig_3}). The Direct Memory Access (DMA) unit retrieves data, including indexes and raw data, from off-chip memory. In addition, instead of fetching high-dimensional data for all neighbors as in~\cite{Peng,Kim}, the pHNSW processor limits the number of irregular accesses to $k$ (for-loop of step \circled{\small{\bf{3}}} in Algorithm~\ref{Algo:P-HNSW}). We use a scratchpad memory (SPM) to store raw data and manage the visit list ($V$-list in Algorithm~\ref{Algo:P-HNSW}). Represented as a 1M-bit state for the SIFT1M dataset, the visit list eliminates redundant computations for previously visited points and requires frequent access.

\begin{table}[h]

\caption{Custom Instruction Overview}
\renewcommand\arraystretch{1.5}
\scriptsize
\begin{center}
\begin{threeparttable}
\vspace{-0.2cm}
\begin{tabular}{|m{0.8cm}<{\centering}|m{1cm}
<{\centering}|m{0.62cm}<{\centering}|m{4.5cm}<{\centering}|}\hline

\textbf{Category} &  \textbf{ISA} &  
\textbf{Cycles} &  \textbf{Description}  \cr\hline
Data & Move& 1 & Move data between registers. \cr\cline{2-4}
Access & DMA & Multi & Read data from off-chip memory.\cr\cline{2-4}
      
& Visit\&Raw& 1 or 2 & Read/Write index or raw data from SPM.  \cr\hline

 &kSort.L &7  & Filter the top-$k$ nearest low-dim distances.\cr\cline{2-4}
 Compute&Min.H &1 &Get the minimum of high-dim distances.\cr\cline{2-4}
 &RMF &8 &Remove indexes from $F$-list.\cr\hline 
 
Control &JMP &1 &Conditional jump operation. \cr\hline

\end{tabular}
    \end{threeparttable}
\label{Table:ISA}
\vspace{-0.3cm}
\end{center}
\end{table}

\subsubsection{Computation Units}

Our pHNSW processor has dedicated computation units to execute computation instructions (Table~\ref{Table:ISA}) during the search process. The \textit{Dist.L} units compute distances in parallel for low-dimensional data (e.g., processing 16 data points simultaneously), matching the SIFT1M graph's configuration where each node links to multiple neighbors. The \textit{kSort.L} unit performs fully parallel sorting of low-dimensional distances, outputting the top-$k$ nearest ones (step \circled{\small{\bf{2}}} in Algorithm~\ref{Algo:P-HNSW}). Fig.~\ref{fig_3}(c) illustrates the \textit{kSort.L} scheme, where all elements are compared simultaneously, and the sorting order is determined by counting $>$ symbols in the comparison matrix. Compared to conventional bubble sort, the fully parallel \textit{kSort.L} design delivers significantly higher search performance. For instance, while bubble sort requires 120 cycles to sort 16 data elements, \textit{kSort.L} completes the same operation in just 7 cycles (94.17\% improvement). The \textit{Dist.H} unit computes distances sequentially for high-dimensional data (step \circled{\small{\bf{3}}} in Algorithm~\ref{Algo:P-HNSW}), while the \textit{Min.H} unit selects the minimum distance(s).

\subsection{Dataflow of pHNSW Processor}
\label{sub:ExecutionOverview}

With the database organized in DRAM and the pHNSW processor design, the dataflow for neighboring search can be summarized in five steps. (1) The pHNSW processor begins the search with the query vector $q$ and a randomly selected entry point $ep$, both in high-dimensional space. (2) The Address Generation Unit (AGU) calculates the address of the neighboring list using the index of $ep$, and provides this address to the DMA, which accesses the neighboring indices and low-dimensional raw data, storing them in the SPM. (3) The Dist.L unit computes the distances between the neighboring raw data and the query $q$ in the low-dimensional space, and the \textit{kSort.L} unit sorts these distances. (4) The AGU and DMA units use the top-$k$ indices from the low-dimensional space to access the corresponding high-dimensional data in DRAM and transfer it to the SPM. (5) As \textit{Dist.H} computes the distances for the top-$k$ points, the \textit{Visit\&Raw} unit checks whether the high-dimensional points have been considered. Throughout the search process, the two \textit{BUS} and two \textit{Move} units manage the transfer of indices and raw data between registers.

\section{Experiment}
\label{sec:Exp}
\subsection{Experiment setup}

\subsubsection{Simulation}
We implement pHNSW processor in Verilog and synthesize the RTL using Synopsys Design Compiler with the TSMC 65nm standard library at 1GHz to obtain the area and power consumption of each component. For the on-chip SPM (128KB), we model area and energy using CACTI 7.0\cite{CACTI}. We use Ramulator \cite{Ramulator} to model DRAM with two standards: 4GB DDR4 and High-Bandwidth Memory (HBM 1.0), with bandwidths of 19.2GB/s and 128GB/s, respectively. We configure the energy consumption at 18.75 pJ/bit for DDR4 and 7 pJ/bit for HBM 1.0.

\subsubsection{Dataset}
We evaluate our P-HNSW design using the SIFT1M dataset~\cite{jegou2010product}. As previously described in Section~\ref{filtersize}, we construct a six-layer search graph \cite{HNSW} and prepare low-dimensional raw data using PCA~\cite{abdi2010principal}. The throughput of our pHNSW is evaluated with the recall of 0.92.

\subsubsection{Baselines}

We run the HNSW algorithm~\cite{HNSW} on a 12th Gen Intel i9-12900H (2.5 GHz) CPU as our primary baseline, denoted as \textit{HNSW-CPU}, and use the reported results from CAGRA~\cite{CAGRA} as our GPU baseline, denoted as \textit{HNSW-GPU}. To evaluate performance gains achieved through algorithmic optimization, we run the proposed pHNSW algorithm (see Section~\ref{P-algo}) on the CPU, denoted as \textit{pHNSW-CPU}. Due to differences in datasets and hardware configurations, a direct comparison with the results reported in~\cite{Peng, song2022accelerating} (molecular and quantum physics datasets) and~\cite{Kim, zhu2023processing,wang2024ndsearch} (million-scale SIFT1B dataset on multi-core or processing-in-memory systems) is not feasible. Our work focuses on single-query search, with plans to scale the \textit{pHNSW} processor to multicore systems for multi-query search in future work.
For a fair comparison, we adapt the instruction execution to our pHNSW processor, creating two variants: (1) \textit{HNSW-Std}, the standard HNSW that processes high-dimensional raw data, and (2) \textit{pHNSW-Sep}, the pHNSW algorithm designed to access low-dimensional raw data stored in a separate database (see \circledd{\small{\bf{4}}} in Fig.~\ref{fig_3}).

\subsection{Area Evaluation}
\label{Area}

Fig.\ref{fig:area} presents the area breakdown of the pHNSW processor, with a total area of $0.739mm^2$. For the on-chip memories, SPM and register files take 37.5\% and 13.9\% area respectively. The register files store temporary data, primarily determined by the data dimensions (e.g., 15 and 128 dimensions, each occupying 4 Bytes) needed for distance calculations and sorting. The \textit{Move} units occupy 23\% of the area due to the extensive use of ports to enable efficient parallel access to high-dimensional data.
The \textit{Dist.L} and \textit{kSort.L} components together occupy 14.0\% area, with \textit{kSort.L} comprising a 16$\times$16 comparator array and four 16-input multiplexers (i.e., for sorting 16 data elements in SIFT1M database).

\begin{figure}[t]
	\centering
\includegraphics[width= 0.9\textwidth]{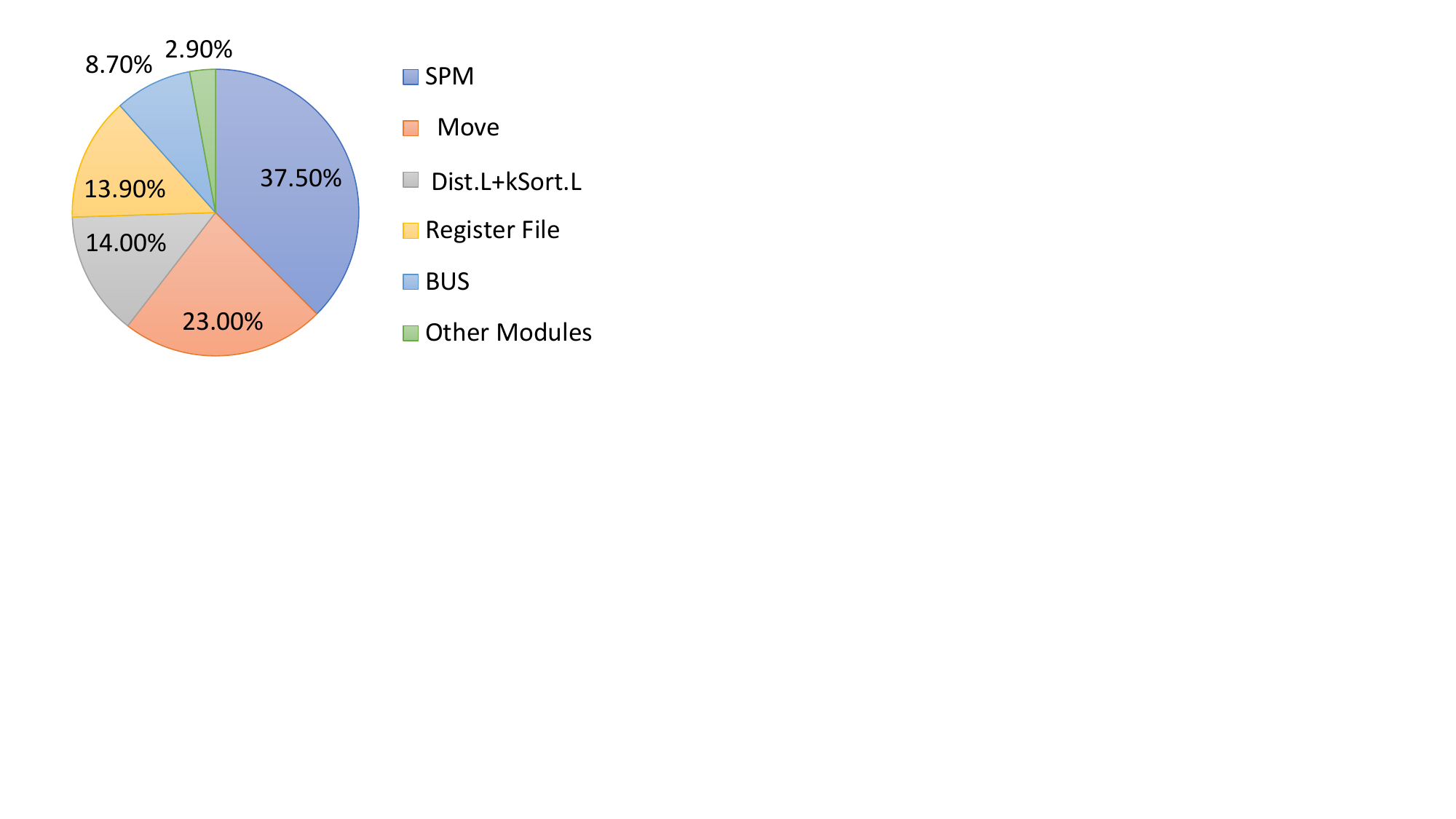}
	\vspace{-5.4cm}
	\caption{Area breakdown of pHNSW processor with a total area of $0.739mm^2$.}
	\label{fig:area}
\end{figure}

\subsection{Search Throughput Evaluation}
\label{QPS}

\begin{table}[h]
\vspace{-0.2cm}
\caption{{Single-query search throughput (QPS) on SIFT1M}}
\renewcommand{\arraystretch}{1.5}
\footnotesize
\begin{center}
\begin{threeparttable}
\vspace{-0.5cm}
\begin{tabular}{|m{1.25cm}<{\centering}|m{1.94cm}
<{\centering}|m{2.1cm}<{\centering}|m{1.84cm}<{\centering}|}\hline

\multirow{2}{*}{QPS }&\textbf{HNSW-CPU}\cite{HNSW}&\textbf{HNSW-GPU}\cite{CAGRA}&\cellcolor{mygray}\textbf{pHNSW-CPU}\cr\cline{2-4} 

\multirow{2}{*}{(Norm.QPS)}&\makecell{\\[-4ex] \\9900.35 
 \\(1)\\[0.5ex]}&\makecell{\\[-4ex]\\$\sim$25000 \\(2.53)}&\makecell{\\\\[-4ex]10467.61 \\(1.06)}\cr\cline{2-4}


&\cellcolor{mygray}\textbf{HNSW-Std}&\cellcolor{mygray}\textbf{pHNSW-Sep}&\cellcolor{mygray}\textbf{pHNSW} (Ours)\cr\hline

DDR4&\makecell{\\[-4ex]\\13290.45 \\(1.74)\\[0.5ex]}&\makecell{\\[-4ex]\\32731.08 \\(3.31)}&\makecell{\\[-4ex]\\143285.14 \\ (14.47)}\cr\hline


HBM&\makecell{\\[-4ex]\\18165.30 \\(1.83)\\[0.5ex]}&\makecell{\\[-4ex]\\77645.78 \\(7.84)}&\makecell{\\[-4ex]\\211579.75\\ (21.37)}\cr\hline

\end{tabular}
    \begin{tablenotes}
    \item *The value in each parenthesis is normalized to the QPS of HNSW-CPU.

    \end{tablenotes}
    \end{threeparttable}
\label{Table:QPS}
\vspace{-0.3cm}
\end{center}
\end{table}

Table~\ref{Table:QPS} compares QPS for single-query searches, leading to several key observations. First, \textit{pHNSW-CPU} shows our algorithm optimization increases QPS by 6\% on the CPU. 
Second, \textit{HNSW-Std} refers to hardware-only optimization, achieving a 1.74×$\sim$1.83× QPS improvement for DDR4 and HBM, while \textit{pHNSW-Sep} involves algorithm and hardware co-optimization without database optimization, resulting in a 3.31×$\sim$7.84× QPS improvement. \textit{HNSW-Std} and \textit{pHNSW-Sep} have similar database structures (\circledd{\small{\bf{2}}} vs. \circledd{\small{\bf{4}}} in Fig.~\ref{fig_3}(a)), with the former operating in high-dimensional space and the latter in low-dimensional space. Third, our proposed \textit{pHNSW} integrates algorithm and hardware co-optimization with database optimization. It achieves QPS improvements of 14.47× and 21.37× for DDR4 and HBM configurations, respectively, compared to \textit{HNSW-CPU}  (\textit{i.e.}, 5.37× and 8.46× improvement vs. \textit{HNSW-GPU}). This is due to regular data access to the sequential storage of low-dimensional neighbor data (\circledd{\small{\bf{3}}} in Fig.\ref{fig_3}(a)), but comes with a 2.92× increase in database memory usage, as explained in Section\ref{sub:dataStorage}.
We consider the trade-off between search throughput and memory usage acceptable for the SIFT1M dataset. For larger datasets like SIFT1B (512GB storage, exceeding typical DRAM capacity), achieving a balance between the \circledd{\small{\bf{3}}} and \circledd{\small{\bf{4}}} in database of Fig.~\ref{fig_3}(a) is crucial and will be further explored in future work.

\subsection{Energy Evaluation of a Single Query Search}

\begin{figure}[h]
    \vspace{-0.3cm}
	\centering
\includegraphics[width= 0.95\textwidth]{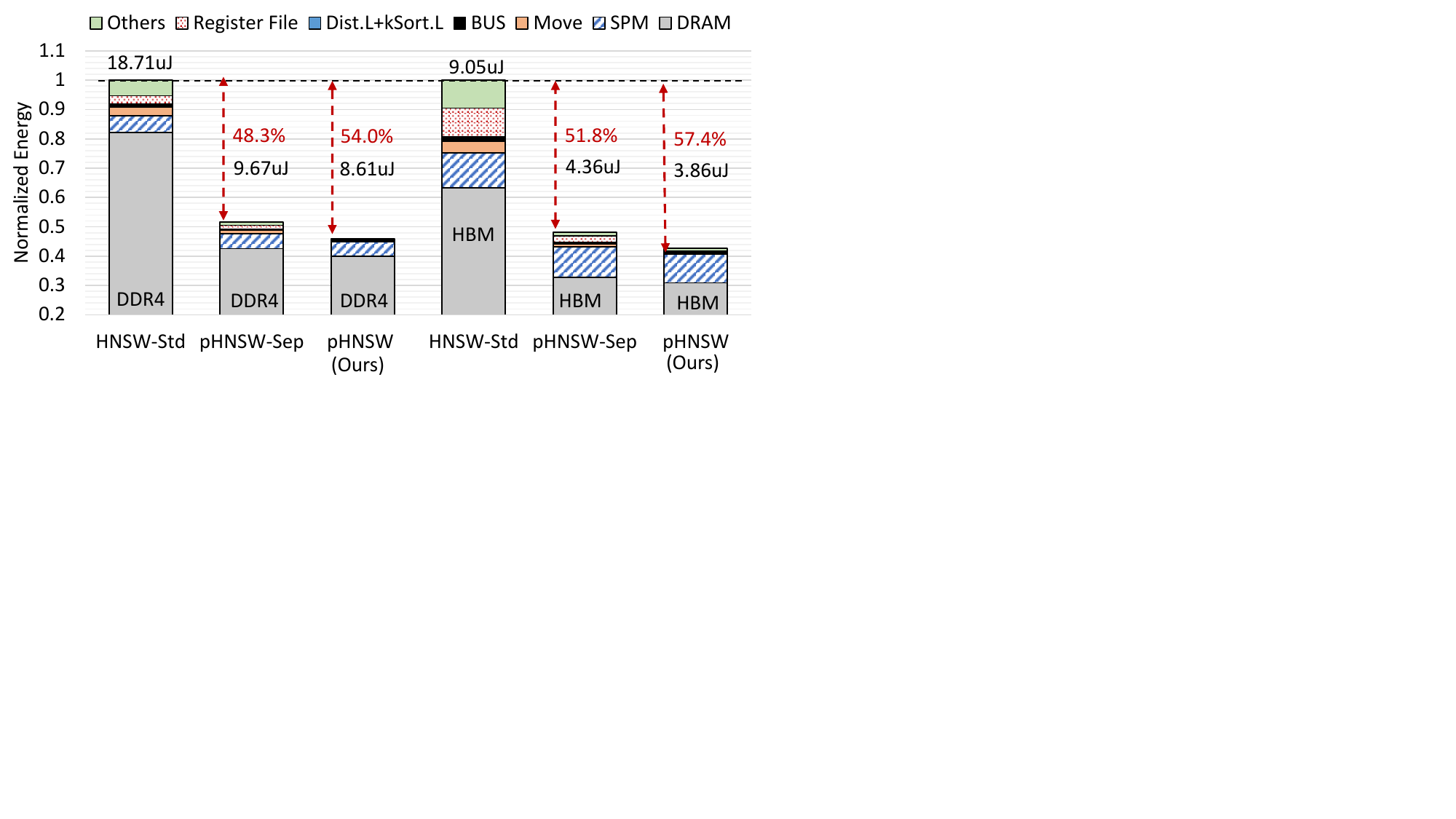}
	\vspace{-5.6cm}
	\caption{Normalized energy of a single query search.}
	\label{fig:Energy}
	\vspace{-0.1cm}
\end{figure}

Fig.\ref{fig:Energy} compares the normalized energy consumption of a single query search (vs. \textit{HNSW-Std}). We have several observations. First, DRAM accesses dominate energy consumption, making up about 82$\sim$87\% and 63$\sim$72\% of the total energy for DDR4 and HBM configurations, respectively.
Second, compared to \textit{HNSW-Std}, \textit{pHNSW-Sep} and \textit{pHNSW} reduce energy consumption by up to 51.8\% and 57.4\%, respectively, showcasing the energy efficiency of our pHNSW design.
Second, although \textit{pHNSW-Sep} and \textit{pHNSW} employ different access patterns (irregular vs. sequential, \circledd{\small{\bf{4}}} vs. \circledd{\small{\bf{3}}} in Fig.~\ref{fig_3}(a)), they retrieve the same amount of data from off-chip memory. Compared to \textit{pHNSW-Sep}, our \textit{pHNSW} design reduces energy consumption by approximately 11\% per query search. 
This reduction is due to the lower latency of \textit{pHNSW}'s regular memory access (resulting in 2.73×$\sim$4.37× higher QPS, in Section~\ref{QPS}), which consequently reduces the energy consumed by other components waiting for data. Third, the dedicated design of the low-dimensional computation module (\textit{Dist.L+kSort.L}) enables high parallelism with
negligible energy consumption ($<$1\%).

\section{Conclusion}
\label{sec:conclusion}
This paper proposes \textit{pHNSW}, an algorithm-hardware co-design of HNSW for high-dimensional search tasks. We optimize the HNSW algorithm using PAC filtering, enabling data access and computation in low-dimensional space while maintaining high search accuracy (recall = 0.92). Additionally, we implement a pHNSW processor with a custom instruction set, leveraging an optimized data organization that stores low-dimensional data to reduce irregular memory accesses. Experiments on the SIFT1M dataset show that the proposed pHNSW processor improves search throughput (QPS) by 14.47×$\sim$21.37× over standard HNSW on CPU and 5.37×$\sim$8.46× on GPU. Moreover, our pHNSW design can achieve up to 57.4\% reduction in energy consumption, compared to the hardware implementation of the standard HNSW in high-dimensional space. 

For future work, since our proposed pHNSW algorithm and hardware design are orthogonal to prior work on multi-core and processing-in-memory (PIM) architectures, we plan to improve search efficiency by scaling the pHNSW processor to multi-core and PIM-based systems. Additionally, we plan to extend our evaluation to the SIFT1B dataset to assess the scalability and performance of the processor on larger and more complex datasets.
A key challenge will be partitioning the billion-scale database into smaller parts while preserving efficient coordination and search performance.

\bibliographystyle{IEEEtran}  
\bibliography{references}     %
\end{document}